\documentclass[osajnl,twocolumn,showpacs,superscriptaddress,10pt]{revtex4-1} %% use 11pt for Applied Optics
\usepackage{amsmath,amssymb,graphicx}
\usepackage{epstopdf}

\begin{document}

%\begin{frontmatter}



%\dochead{Optics Comm}
%% e.g. \dochead{17th International Conference on Dynamical Processes in Excited States of Solids}

\title{Ray transfer matrix for a spiral phase plate}

%%use optional labels to link authors explicitly to addresses:
\author{M. Eggleston}
\affiliation{School of Physics and Astronomy, Rochester Institute of Technology, 84 Lomb Memorial Drive,
Rochester, NY 14623}
\author{T. Godat}
\affiliation{School of Physics and Astronomy, Rochester Institute of Technology, 84 Lomb Memorial Drive,
Rochester, NY 14623}
\author{E. Munro}
\affiliation{School of Mathematical Sciences, Rochester Institute of Technology, 85 Lomb Memorial Drive,
Rochester, NY 14623}
\author{M. A. Alonso}
\affiliation{The Institute of Optics, University of Rochester, Rochester, NY 14627}
\author{H. Shi}
\affiliation{School of Physics and Astronomy, Rochester Institute of Technology, 84 Lomb Memorial Drive,
Rochester, NY 14623}
\author{M. Bhattacharya}
\affiliation{School of Physics and Astronomy, Rochester Institute of Technology, 84 Lomb Memorial Drive,
Rochester, NY 14623}

\begin{abstract}
We present a ray transfer matrix for a spiral phase plate. Using this matrix, we determine the
stability of an optical resonator made of two spiral phase plates, and trace stable ray orbits
in the resonator. Our results should be relevant to laser physics, optical micromanipulation,
quantum information and optomechanics.
\end{abstract}

%\ocis{(080.4865) Optical vortices; (140.4780) Optical resonators; (260.6042)
%Singular optics; (080.3095) Inhomogeneous elements in optical systems.}

%\begin{keyword}



%Spiral phase plates, orbital angular momentum, ray transfer matrix

%\end{keyword}

%\end{frontmatter}

% \linenumbers

\maketitle

\section{Introduction}
\label{1}
The spiral phase plate is an important element in modern optics because it can impart well-defined orbital angular
momentum to any photon which it reflects or transmits \cite{AllenBook}. Spiral phase plates are widely employed as
mode selectors, in active cavities \cite{DavidsonReview}, as etalons \cite{Rumala2013} and in free space
\cite{Beijersbergen1994}; in the rotation of microparticles \cite{Lee2004} and nanomechanical elements
\cite{Bhattacharya2007,Isart2010}; in microscopy \cite{Marte2008}; in astronomy
\cite{Swartzlander2005}; and in tests of quantum mechanics \cite{Woerdman2005}.

To the best of our knowledge, a ray transfer matrix has not yet been presented for the spiral phase element. Such a
treatment is desirable, since, as is well known, the ray matrix provides an insightful and useful description
of most basic optical elements. Thus, ray matrix analyses of mirrors and lenses, of optical resonator stability, and
of fiber waveguiding can be found in several optics textbooks \cite{SiegmanBook,EberlyBook}. There is also a practical
need for a spiral phase plate ray matrix, so that ray optics can be extended to systems that integrate spiral phase
plates with standard optical components \cite{DavidsonReview,Swartzlander2005}, without necessitating recourse to the
more complicated wave-optical diffraction theory. A description of the effect of a spiral phase plate on light rays
has been presented earlier \cite{Padgett1996}, highlighting the role of orbital angular momentum transfer, but without
reference to a ray transfer matrix.

In this article we derive a ray matrix which describes the reflection of light rays from a spiral phase plate
carrying an azimuthal gradient as well as a radial curvature. We use the matrix to find the stability condition for
a resonator made of two spiral phase plates. This spiral resonator is the rotational analog of the standard spherical
mirror Fabry-Perot \cite{SiegmanBook}, and has been discussed earlier in the literature in the context of laser physics
\cite{PaxtonBook,DavidsonReview} and optomechanics \cite{Bhattacharya2007}. We provide a simple analytical criterion for
resonator stability, which reduces to the standard expression for a spherical mirror cavity in the absence of azimuthal
structure on the phase plates. We also use the derived matrix to trace stable ray orbits in the spiral resonator.
\section{The spiral phase plate}
\label{sec:SPP}
\begin{figure}[]
\centerline{\includegraphics[width=1.0\columnwidth]{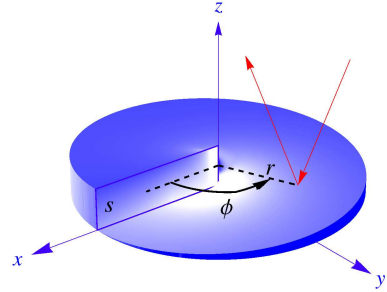}}
\caption{The spiral phase plate discussed in this work. An incident ray is shown being reflected by the plate.}
\label{fig:F1}
\end{figure}

A spiral phase plate is an optical element whose thickness increases linearly with the azimuthal angle, as shown in
Fig.~\ref{fig:F1}.
The plate is characterized by the step discontinuity of height $s$, and the local pitch angle $\alpha(r)$ at
the radial coordinate $r$ given by \cite{Beijersbergen1994}
\begin{equation}
\label{eq:pitch}
\tan \alpha(r) = \frac{s}{2\pi r}.
\end{equation}
For the purpose of this study, the center of the plate will be placed on the optical axis. Further, only the paraxial
approximation will be considered, in which light rays make small angles to the optic axis \cite{SiegmanBook}. Thus
we will assume that $\tan\alpha(r)\simeq \alpha(r)\simeq s/2\pi r.$ In turn, this implies that $r \gg s/2\pi$. In
other words, we will only consider rays impinging far away from the center of the spiral phase plate.

In addition to the azimuthal gradient described by Eq.~(\ref{eq:pitch}) we will also assume that the spiral phase plate
has a radial gradient like an ordinary (concave) spherical mirror, characterized by a radius of curvature $R$. The
presence of this curvature is essential for ensuring the stability of the spiral phase plate resonator described
below, and also allows us to relate our analysis to the standard results of ray optics theory for $s=0.$
\section{Reflection ray matrix}
\label{sec:RayM}
To obtain the ray transfer matrix for the spiral phase plate, we begin with a wavefront aberration approach
\cite{Jeong2005}. Since the plate presents a nonorthogonal optical system \cite{SiegmanBook}, the ray is
represented by a four-dimensional vector. We define this vector as $(r,dr/dz,\phi,rd\phi/dz),$ with the first two
entries denoting the radial position and inclination respectively with respect to the optic $(z)$ axis, and the last
two signifying the corresponding azimuthal quantities. The ray transfer matrix is given by \cite{Jeong2005}
\begin{equation}
\label{eq:Mcs}
M_{c}(s,R)=
\left(\begin{array}{cc}
M_{r} & 0 \\
0 & M_{t}\
\end{array}\right),
\end{equation}
where
\begin{equation}
\label{eq:rad}
M_{r}=
\left(\begin{array}{cc}
1 & 0 \\
-\frac{1}{r_{1}}\left.\frac{\partial W(r,\phi)}{\partial r}\right|_{\textbf{r}=\textbf{r}_{1}} & 1\
\end{array}\right),
\end{equation}
is the radial matrix and
\begin{equation}
\label{eq:tan}
M_t=
\left(\begin{array}{cc}
1 & 0 \\
-\frac{1}{\phi_{1}}\left.\frac{\partial W(r,\phi)}{r\partial \phi}\right|_{\textbf{r}=\textbf{r}_{1}} &1\
\end{array}\right),
\end{equation}
is the tangential matrix. In Eqs.~(\ref{eq:rad})-(\ref{eq:tan}), $W(r,\phi)$ is the wave front aberration for
an optical element, defined as the deviation from an arbitrary reference plane perpendicular to the optic axis \cite{Jeong2005}.
For the spiral phase plate, we find
\begin{equation}
\label{eq:wab}
W(r,\phi)=\frac{r^{2}}{R}+s\left(1-\frac{\phi}{2\pi}\right).
\end{equation}
Using Eqs.~(\ref{eq:Mcs})-(\ref{eq:wab}), we arrive at
\begin{equation}
\label{eq:cylcoord}
M_{c}(s,R) =
\left(\begin{array}{cccc}
1 & 0 & 0 & 0 \\
-\frac{2}{R} & 1 & 0 & 0 \\
0 & 0 & 1 & 0 \\
0 & 0 & \frac{s}{\pi r_{1}\phi_{1}} & 1 \
\end{array}\right).
\end{equation}
Although the wavefront aberration formalism has been used since it is general and compact, the
matrix $M_{c}(s,R)$ can be readily derived by using geometric ray diagrams and Snell's law for reflection in the
$r-z$ and $\phi-z$ planes respectively (see Fig.~\ref{fig:F1}), in the usual textbook manner.

It is important to note that unlike most ray matrices described in textbooks, the matrix $M_{c}(s,R)$ is inhomogeneous,
meaning that it depends on the ray coordinates $r_{1}$ and $\phi_{1}$. This dependence underlines the fact that the ray
reflection from the spiral phase plate is inherently nonlinear, similar to the case of optical elements with coma or
spherical aberration, for example \cite{Jeong2005}. Inhomogeneous ray transfer matrices can nevertheless be useful for
analyzing resonator stability, as shown for the Bessel-Gauss resonator \cite{Cerda2003}, and for accurate ray-tracing
\cite{Jeong2005}. Similarly, we analyze below the stability of a spiral phase plate resonator using the matrix
$M_{c}(s,R)$ and trace some of the stable ray orbits.

For the following analysis, it is convenient to transform the ray transfer matrix of Eq.~(\ref{eq:cylcoord}) to Cartesian
coordinates. Using the well known relations
\begin{eqnarray}
\label{eq:cylCart}
x&=r\cos\phi, & y=r\sin\phi,\\
r& = \sqrt{x^{2}+y^{2}}, & \phi = \tan^{-1}\left(\frac{y}{x}\right),
\end{eqnarray}
we can calculate the derivatives
\begin{eqnarray}
\label{eq:diff}
\frac{dr}{dz}&=\frac{x\frac{dx}{dz}+y\frac{dy}{dz}}{\sqrt{x^{2}+y^{2}}},&
\frac{d \phi}{dz}=\frac{x\frac{dy}{dz}-y\frac{dx}{dz}}{x^{2}+y^{2}}.
\end{eqnarray}
Using Eqs.~(\ref{eq:cylCart})-(\ref{eq:diff}), we transform Eq.~(\ref{eq:cylcoord}) to read
\begin{equation}
\label{eq:CartM}
M(s,R) =
\left(\begin{array}{cccc}
1 & 0 & 0 & 0 \\
-\frac{2}{R} & 1 & \frac{s}{2\pi r_{1}^{2}} & 0 \\
0 & 0 & 1 & 0 \\
-\frac{s}{2\pi r_{1}^{2}} & 0 & -\frac{2}{R} & 1 \
\end{array}\right),
\end{equation}
where we have used $x_{1}=x_{2},y_{1}=y_{2}, r_{1}=r_{2},$ and $r_{1}^{2}=x_{1}^{2}+y_{1}^{2},$
etc.

The determinant of $M(s,R)$ is one, indicating that the plate is lossless. Note however that, unlike the usual
non-orthogonal ray transfer matrices in the literature, the matrix in Eq.~(\ref{eq:CartM}) is not symplectic,
namely, it does not satisfy the condition
\begin{equation}
\label{eq:symplectic}
\sigma\cdot M^{\rm T}\cdot\sigma\cdot M=-I,
\end{equation}
for $M=M(s,R)$, where $I$ is the identity matrix and, for the ray-vector ordering used here,
\begin{equation}
\label{eq:sigma}
\sigma =
\left(\begin{array}{cccc}
0 & -1 & 0 & 0 \\
1 &0 & 0 & 0 \\
0 & 0 & 0 & -1 \\
0 & 0 & 1 & 0 \
\end{array}\right).
\end{equation}
The symplecticity condition guarantees not only the conservation of \'etendue, but also ensures that
any incident two-parameter normal congruence of rays (i.e., rays that are normal to a family of wavefronts) emerges also as a normal congruence. However, this condition as expressed in Eq.~(\ref{eq:symplectic}) is only valid for homogeneous matrices $M$, that is, for
\textit{linear} ray mappings. For more general ray mappings it is not $M(s,R)$ that must be symplectic, but the Jacobian
matrix of  the final ray parameters $(x_2,p_2,y_2,q_2)=M(s,R)\cdot(x_1,p_1,y_1,q_1)$ with respect to the initial ray
parameters $(x_1,p_1,y_1,q_1)$, where $p=dx/dz$ and $q=dy/dz$. (Note that, for a homogeneous ray matrix $M$, this
Jacobian matrix does reduce to $M$ itself.) The Jacobian matrix is symplectic if the following conditions are satisfied
\cite{Minano1984}:
\begin{eqnarray}
\frac{\partial x_2}{\partial u}\frac{\partial p_2}{\partial v}-\frac{\partial x_2}{\partial v}\frac{\partial p_2}{\partial u}+\frac{\partial y_2}{\partial u}\frac{\partial q_2}{\partial v}-\frac{\partial y_2}{\partial v}\frac{\partial q_2}{\partial u}\nonumber
\\
=\left\{\begin{array}{cc}1,&u,v=x_1,p_1\,{\rm or}\,y_1,p_1,\\0,&u,v=x_1,y_1\,{\rm or}\,x_1,q_1\,{\rm or}\,y_1,p_1\,{\rm or}\,\,p_1,q_1.\end{array}\right.
\end{eqnarray}
It can be readily verified that these conditions are satisfied for the matrix $M(s,R)$ in Eq.~(\ref{eq:CartM}).
\section{Spiral phase plate resonator}
\label{sec:TwoSPP}
\begin{figure}[]
\centerline{\includegraphics[width=1.0\columnwidth]{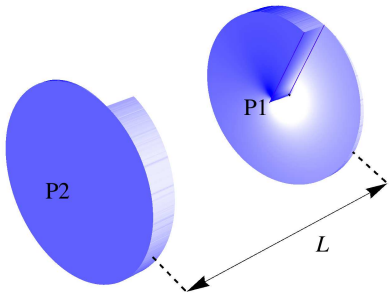}}
\caption{A spiral resonator with two identical plates separated by a distance $L$. The steps of the plates have been aligned.}
\label{fig:F2}
\end{figure}

We now consider, as an application of the transfer matrix of Eq.~(\ref{eq:CartM}), the stability of a
resonator made of two spiral phase plates separated by a distance $L$, as shown in Fig.~\ref{fig:F2}.
For simplicity, we assume that the two spiral phase plates are identical.

To apply the inhomogeneous ray matrix of Eq.~(\ref{eq:CartM}) to the spiral phase plate resonator, we consider a
situation where $M(s,R)$ becomes effectively homogeneous. This can be accomplished if $r_{1}$ assumes a fixed value $r$, i.e.,
if the rays always strike the two plates at the same distance from the optic axis. A self-consistent and stable ray optics
solution can indeed be found for this case, as we now show. To calculate the resonator stability, we begin with a light
ray just to the left of the plate $P1$, and propagate it towards $P2$ by a distance $L$ using the matrix
\cite{EberlyBook}
\begin{equation}
\label{eq:fspacel}
M(L)=
\left(\begin{array}{cccc}
1 & L & 0 & 0 \\
0 & 1 & 0 & 0 \\
0 & 0 & 1 & L \\
0 & 0 & 0 & 1\
\end{array}\right).
\end{equation}
Subsequent reflection of the ray from the plate $P2$ at the radial point $r$ is modeled by the matrix $M(s,R)$
[see Eq.~(\ref{eq:CartM})].
The light ray then travels back to plate $P1$, again propagated by the matrix $M(L).$ Finally, the ray is reflected
at the radial coordinate $r$, by plate $P1$. It is therefore multiplied by $M(-s, R),$ which can be obtained from
Eq.~(\ref{eq:CartM}). Note that the sign of $s$ is opposite for $P1$ and $P2$, although the two plates have the same winding,
because they face each other. The resulting round trip matrix is defined by
\begin{equation}
\label{eq:Mrt}
M_{T}=M(-s,R)\cdot M(L)\cdot M(s,R)\cdot M(L),
\end{equation}
which has not been presented explicitly as it is rather complicated in structure and can be found readily
using a symbolic computation package, such as \textit{Mathematica}.

To arrive at the stability condition we solve the characteristic polynomial of $M_{T}$,
\begin{equation}
P(\lambda)=|M_{T}-\lambda I|=0,
\end{equation}
where $I$ is the unit matrix, for the eigenvalues $\lambda$. We find the four eigenvalues to be a twofold degenerate
complex conjugate pair
\begin{equation}
\lambda_{\pm}=e^{\pm i\theta},
\end{equation}
where
\begin{equation}
\label{eq:cos}
\cos \theta=1-\frac{4L}{R}+2\left(\frac{L}{R}\right)^{2}+\frac{1}{2}\left(\frac{sL}{\pi r^{2}}\right)^{2}.
\end{equation}
We now make several observations about Eq.~(\ref{eq:cos}). First, if there is no winding on the plate, i.e. for $s=0,
$ Eq.~(\ref{eq:cos}) recovers the textbook result for a spherical mirror cavity \cite{EberlyBook}. In fact, given
the complexity of the round trip matrix $M_{T}$ of Eq.~(\ref{eq:Mrt}), we find it remarkable that the presence of
winding results in only a single additional term. Second, we note that Eq.~(\ref{eq:cos}) does not change if the
handedness is changed from $s \rightarrow -s$, as might be expected on grounds of symmetry. Third, we observe that
for physically realizable parameters, $\cos\theta$ in Eq.~(\ref{eq:cos}) can be real, which is a precondition for
the stability of ray trajectories in a resonator.

We now quantify the criterion for stability, which is usually written as \cite{EberlyBook}
\begin{equation}
\label{eq:costab}
-1 \leq \cos\theta \leq 1.
\end{equation}
Using Eq.~(\ref{eq:cos}), the left inequality in Eq.~(\ref{eq:costab}) can be shown to lead to the relation
\begin{equation}
\left(1-\frac{L}{R}\right)^{2}+\left(\frac{sL}{2\pi r^{2}}\right)^{2} \geq 0,
\end{equation}
which is always satisfied since the left hand side of the inequality is the sum of squares of two real numbers.
The right inequality in Eq.~(\ref{eq:costab}) can similarly be simplified to
\begin{equation}
\label{eq:rstab}
L \leq \frac{2R}{1+\left(\frac{sR}{2\pi r^{2}}\right)^{2}}.
\end{equation}
It is revealing to verify the physical consistency of this simple stability condition in various parametric
limits. The standard spherical cavity result is recovered in
the absence of winding $(s=0)$. For a fixed length $L$, the presence of winding $(s \neq 0)$, destabilizes the
cavity if the step size is large $(s\rightarrow \infty)$. For a fixed $L$ and nonzero winding, the cavity becomes
unstable both for a small radius of curvature $(R \rightarrow 0)$ as well as in the limit of a ``plane" phase
plate $(R\rightarrow \infty).$ The resonator is also destabilized if the ray radius is small $(r\rightarrow 0),$
while the spherical mirror resonator stability condition $(L \leq 2R)$ is recovered for a large ray radius
$(r\rightarrow \infty).$ For the parameters $R=10$cm, $s=10\mu$m, and $r=0.5$mm, we find the denominator in
Eq.~(\ref{eq:rstab}) to be 1.4, which should be a measurable shift in the stability boundary from $L=2R$.
\section{Ray tracing of stable orbits}
\begin{figure}[!]
\centerline{\includegraphics[width=1.0\columnwidth]{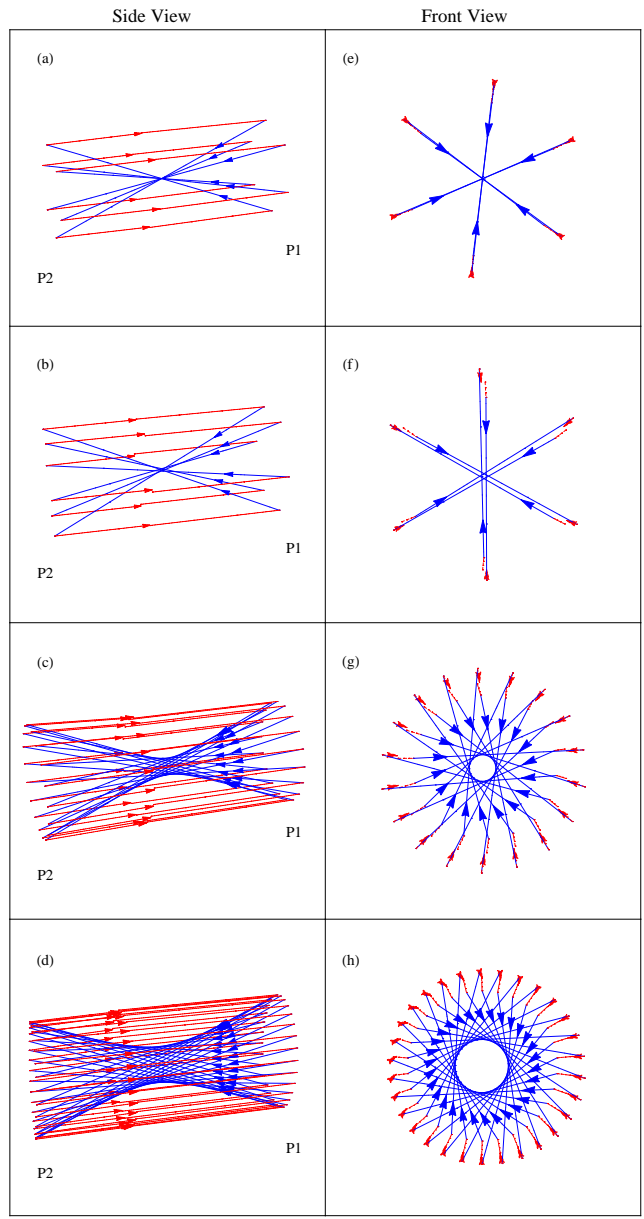}}
\caption{Stable ray orbits that can occupy the spiral plate resonator of Fig~\ref{fig:F2}. The plates $P1$
and $P2$ are not shown for clarity. Solid arrows denote rays traveling from $P1$ to $P2$ and dotted arrows rays traveling
from $P2$ to $P1$. Figures (a), (b), (c) and (d) are the side views showing the increasing ``twist" in the
ray bundle as the step height $s$ is increased. Figures (e), (f), (g), and (h) are the corresponding front views,
i.e. along the optic axis, showing how the central region is increasingly avoided by the rays as $s$ becomes larger.
This dark region is consistent with the existence of a vortex.
(a) $s=0, r = 12$mm, $R = 59$mm, $L = 60$mm, number of rays = 6, $N$(number of round-trips)$=2$.
This corresponds to the case where the plate is simply a concave mirror, and the rays focus at the center of the
cavity. (b) $s = 0.5$mm, $r = 14$mm, $R = 59$mm, $L = 62$mm, number of rays = 6, $N = 2$. (c) $ s = 2.5$mm, $r = 12$mm,
$R = 59$mm, $L = 62$mm, number of rays $= 18, N=9$. (d)$s=5$mm, $r=12$mm, $R=59$mm, $L=55$mm, number of rays = 30,
$N=5$.}
\label{fig:F3}
\end{figure}

Shown in Fig.~\ref{fig:F3}
are stable ray orbits in the spiral plate resonator of Fig.~\ref{fig:F2}. For clarity, the spiral plates have not been
drawn. The ray orbits were found as follows. A ray parallel to the optic axis was initially assumed to be incident
on plate $P1$. The point of intersection of the ray with $P1$ was found numerically taking into account the radial
as well as azimuthal gradients of the plate. The matrix $M(s,R)$ [Eq.~(\ref{eq:CartM})] was then used to account
for the reflection of these rays. The rays were then propagated to $P2$ where a similar procedure
was followed. Closed orbits were found by choosing the resonator parameters obeying the stability condition
of Eq.~(\ref{eq:rstab}). The number of resonator round trips required for the ray orbit to close upon
itself is given by the smallest integer $N$ such that $N(2\pi/\theta)$ is an integer, where $\theta$
is defined in Eq.~(\ref{eq:cos}). Specific cases are discussed in the caption of Fig.~\ref{fig:F3}.
\section{Conclusions}
\label{4}
We have presented a ray transfer matrix for a spiral phase plate. We have used this matrix to derive a simple
analytical stability condition for a Fabry-Perot resonator made of two such spiral phase plates. We have also
presented traces of stable ray orbits. We have only treated the case where the rays are incident at the same distance from the optic
axis on each plate. More general configurations, allowing for different ray radii at the two spiral phase plates,
or for two plates with different radii of curvature, or for rays which strike each plate at more than one radius,
will be investigated in the future. Also, we have restricted our treatment to a ray picture, and thus not considered
any quantized angular momentum or discrete vorticity, as these require some discussion of the wave model of light.
Future work will be aimed at exploring this wave-optical nature of the resonator beams, including diffractive losses.
We expect our present results to be useful to scientists working on laser physics, optical micromanipulation, quantum
information and optomechanics.
\section{Acknowledgements}
We are grateful to G. Swartzlander, S. Preble, E. Hach, N. Davidson and P. K. Lam for useful discussions.
We also thank the Research Corporation of Science Advancement for support. M.E. and E.M. are grateful to the
Rochester Institute of Technology for an Undergraduate Summer Research Award and a Dean's Research Initiation
Grant, respectively. M.A.A. acknowledges support from the National Science Foundation (PHY-1068325).



%% References


\bibliographystyle{elsarticle-num}



\end{document}